\documentclass[journal]{IEEEtran}
\usepackage{cite}
\usepackage{amsmath,amssymb,amsfonts,amsthm}
\usepackage{algorithmic}
\usepackage{graphicx}
\usepackage{textcomp}
\usepackage{caption}
\usepackage{subfigure}
\usepackage{booktabs}
\usepackage[colorlinks=true, allcolors=blue]{hyperref}
\usepackage[linesnumbered,ruled,vlined]{algorithm2e}
\usepackage{wrapfig}
\usepackage{cite}
\usepackage{xr}
\usepackage{hyperref}
\usepackage{bm}
\usepackage[normalem]{ulem}

\makeatletter
\newcommand*{\addFileDependency}[1]{
  \typeout{(#1)}
  \@addtofilelist{#1}
  \IfFileExists{#1}{}{\typeout{No file #1.}}
}
\makeatother
\newcommand*{\myexternaldocument}[1]{
    \externaldocument{#1}
    \addFileDependency{#1.tex}
    \addFileDependency{#1.aux}
}
\myexternaldocument{Tdenoisesupplemental}

\usepackage{adjustbox}
\usepackage{setspace}
\usepackage{hyperref}
\usepackage{multirow}
\usepackage[switch,columnwise]{lineno}
\usepackage{chapterbib}
\graphicspath{ {./figure/} }
\newcommand{\etal}{\textit{et al}.}

\SetKwInput{KwInput}{Input}                % Set the Input
\SetKwInput{KwOutput}{Output}              % set the Output

\def\BibTeX{{\rm B\kern-.05em{\sc i\kern-.025em b}\kern-.08em
    T\kern-.1667em\lower.7ex\hbox{E}\kern-.125emX}}
\begin{document}

%\title{Optimizing Binary Signal Detection Task Performance 
%Using a Hybrid Loss-enabled Model Training Method for Image Denoising}

%\title{Investigation of a task-informed hybrid loss approach for learning-based image denoising}
\title{On the impact of incorporating task-information in learning-based image denoising}

\author{Kaiyan Li, \IEEEmembership{Student Member, IEEE}, Hua Li, and Mark A. Anastasio \IEEEmembership{Senior Member, IEEE}
\thanks{
Kaiyan Li is with the Department of Bioengineering, University
	of Illinois Urbana-Champaign, Urbana, IL, 61801 USA (Email:
	kaiyanl2@illinois.edu).
Hua Li is with the Department of Radiation Oncology, Washington University School of Medicine in St. Louis, Saint Louis, MO.
She is also with the Department of Bioengineering, University of Illinois Urbana-Champaign, Urbana, IL 61801 USA. (Email:li.hua@wustl.edu \& huali19@illinois.edu).
Mark A. Anastasio is with the Department of Bioengineering, University
	of Illinois Urbana-Champaign, Urbana, IL, 61801 USA (Email:
	maa@illinois.edu).
This work was supported in part by NIH awards R01EB020604, R01EB023045, R01NS102213, and R01CA233873, Cancer Center at Illinois seed grant, Jump ARCHES award, and DOD Award E01 W81XWH-21-1-0062.
(Corresponding authors: Mark A. Anastasio \& Hua Li).
}
\vspace{-0.6cm}
}
\maketitle
% \modulolinenumbers[5]
% \linenumbers

\begin{abstract}
A variety of deep neural network (DNN)-based image denoising methods have been proposed for use with medical images. These methods are typically trained by minimizing loss functions that quantify a distance between the denoised image, or a transformed version of it, and the defined target image (e.g., a noise-free or low-noise image). They have demonstrated high performance in terms of traditional image quality metrics such as root mean square error (RMSE), structural similarity index measure (SSIM), or peak signal-to-noise ratio (PSNR). However, it has been reported recently that such denoising methods may not always improve objective measures of image quality. 
In this work, a task-informed  DNN-based image denoising method was established and systematically evaluated.  
A transfer learning approach was employed, in which the DNN is first pre-trained by use of a conventional (non-task-informed) loss function and subsequently fine-tuned by use of the hybrid loss that includes a task-component. The task-component was designed to measure the performance of a numerical observer (NO) on a signal detection task.
The impact of network depth and constraining the fine-tuning to specific layers of the DNN was explored.
The task-informed training method was investigated in a stylized low-dose X-ray computed tomography (CT) denoising study for which binary signal detection tasks under signal-known-statistically (SKS) with background-known-statistically (BKS) conditions were considered. 
The impact of changing the specified task at inference time to be different from that employed for model training, a phenomenon we refer to as ``task-shift", was also investigated.
The presented results indicate that the task-informed training method can improve observer performance while providing control over the trade off between traditional and task-based measures of image quality.

\end{abstract}


\section{Introduction}
\label{sec:purpose}  % \label{} allows reference to this section
%Denoising is a classical and commonly employed image restoration problem in medical imaging applications~\cite{manduca2009projection,lin2001improving,le2013denoising}.  
The development of image denoising methods for medical imaging applications based on deep neural networks (DNNs) remains an active area of research  ~\cite{zhang2017beyond,gong2018pet,you2019denoising}.
Although  learning-based image denoising methods, by conventional design, can improve traditional image quality (IQ) measures such as root mean square error (RMSE) and structural similarity index measure (SSIM), it is well-known that such measures may not always correlate with objective task-based IQ measures \cite{barrett2013foundations,li2021assessing,yu2020ai,li2021task}. Here and throughout this article, a ``task'' denotes an image-based inference to be performed by a human or numerical observer. 
This is because the loss functions that are commonly employed to train such methods do not explicitly take into account the intended task that is to be performed by use of the resulting images.
For example, Yu \emph{et al.} \cite{yu2020ai} demonstrated that task-based metrics were not consistent with traditional IQ metrics in a  study of DNN-based image denoising related to nuclear medicine imaging. 
%This clearly motivated the need for objective evaluation of artificial intelligence (AI)-based methods on clinically relevant tasks \cite{yu2020ai}.
Likewise, Li \emph{et al.} \cite{li2021assessing} reported similar findings and systematically
investigated task-related information loss induced by DNN-based denoising methods under different conditions. 
Task-information loss has also been studied within the context of the learning-based single-image super-resolution problem \cite{zhang2021impact}.
%\cite{li2021assessing}. 

%Most of the available DNN-based image restoration methods are established by use of a loss function defined by a prescribed distance between degraded and target images, or some transformed versions of the images.
%The visibility of important structural details can be compromised or reduced \cite{yang2018low} in the denoised images during the training process, which may degrade the task performance.

To enhance the utility of an image produced by use of a learning-based method, information regarding a task can be naturally incorporated into the  training procedure  \cite{adler2021task,zhang2021task,ongie2022optimizing,han2021low}.
%For example, 
A variety of task-informed methods employ a hybrid loss  comprised of a conventional and a task-based loss component.
For an image reconstruction problem, Adler \emph{et al.} \cite{adler2021task} proposed such an approach to establish a learned reconstruction operator.
%In that method,  tasks that can be encoded as a trainable neural network acting on model parameters and a decision rule for model training. 
Similarly, Ongie \emph{et al}. \cite{ongie2022optimizing} designed a low-dose CT reconstruction framework to enhance the detectability of signals.
For enhancing the utility of denoised images for segmentation tasks, Zhang \emph{et al.} \cite{zhang2021task} proposed a task-informed low-dose CT denoising framework that employed a hybrid loss that incorporated the dice score loss. 
In a different approach that did not employ the hybrid loss strategy, Han \emph{et al.} \cite{han2021low} proposed a perceptual loss-based denoising method.
%That method employs a CNN-based observer as the feature extractor to replace the traditional VGG network \cite{simonyan2014very}.
%The studies described above show promising results in preserving task-based information of the trained network. 
%Preliminary result analysis has been reported in some of those methods.

While these studies provided valuable insights into the potential of learning-based task-informed image restoration methods, this line of research is relatively new and underdeveloped.  
Improved utility of the restored image for the specified task generally comes at the cost of degraded task-agnostic measures of image quality, and understanding this complicated trade-off within the context of a specific problem is important.
% A very common family of tasks in medical imaging applications addresses signal detection \cite{sharp1996report}, but a systematic investigation of task-informed restoration methods for such tasks has not been reported \cite{adler2021task,zhang2021task,ongie2022optimizing,han2021low}.
{While} task-related information has been incorporated into loss function designs, the use of transfer learning coupled with constraints on how {such} information is utilized during model fine-tuning remains unexplored. This is potentially important because previous studies have reported that task-related information may be primarily lost by the deeper layers of a DNN for certain applications \cite{li2021assessing}.

Another critical issue that is specifically relevant to task-informed learning-based methods for image restoration or formation relates to generalization performance
with respect to the task. Tasks in medical imaging applications are generally complicated and can be difficult to comprehensively specify, either analytically or implicitly via specification of a collection of acquired images.  For example, a signal detection task requires the specification of the signal to be detected and the background in which it is embedded. Both of these quantities are generally stochastic in nature and will vary with subject and disease state in the specified cohort.
When a task-informed image restoration method is trained with consideration of a specified detection task, {by design,} it is anticipated that the resulting images will possess enhanced utility for performing that particular task. However, at inference time, the characteristics of the signal or background may differ from those modelled in the original task.
This is a  phenomenon that we refer to as “task-shift”, indicating that \textit{source tasks} (used for training) are different from  \textit{target tasks} (used for inference) \cite{pan2009survey}.
% which can be viewed as a particular form of domain shift. 
Assessing the robustness of a task-informed image restoration method to task-shift is essential to understand its potential suitability for clinical translation.
To date, this topic has not been explored within the context of task-informed learning-based image restoration or reconstruction methods.

To begin addressing these issues,  in this work a task-informed DNN-based image denoising method is established with consideration of  
several signal-known-statistically (SKS) with background-known-statistically (BKS) binary signal detection tasks of varying difficulty. The method is systematically and objectively assessed by use of a stylized low-dose X-ray computed tomography (CT) denoising testbed 
that involves detecting lesions in denoised low-dose CT lung images.
A transfer learning approach is employed, in which {a} DNN is pre-trained by use of a conventional (non-task-informed) loss function and subsequently fine-tuned by use of a hybrid loss that includes a task-component. The task-component  is designed to measure the performance of a numerical observer (NO) on the detection task.
The impact of network depth and constraining the fine-tuning to specific layers of the DNN is explored.
Additionally, the impact of task-shift is investigated to assess the robustness of the approach.
It is expected that the study findings will provide insights into the design and assessment of other task-informed learning-based methods for image restoration and image reconstruction.

The remainder of the paper is organized as follows. 
Section~\ref{sec:background} describes the necessary background on DNN-based image denoising, signal detection tasks, and numerical observers. The  task-informed training method is described in Section~\ref{sec:methods}.
The numerical studies and the corresponding evaluations of the task-informed training method are provided in Sections~\ref{sec:numerical_studies} and~\ref{sec:results}, respectively.
Finally, the article provides a discussion of the key findings in Section~\ref{sec:summary}.

% demonstrate the trade off between traditional and task-based measures. The 

% For the case where the signal location is random, a single layer neural network-based numerical observer (SLNN-NO) \cite{zhou2020approximating} is employed to demonstrate the trade off between traditional and task-based measures. The 

% The performance of the SLNN-NO acting on the images denoised by networks that are trained by use of a hybrid loss function consisting of a conventional and a task-based loss component is quantified via receiver operating characteristic (ROC) analysis. 
% In addition, the impact of the number of trainable layers specified in the denoising network on the SLNN-NO performance is assessed to gain insights into the contribution from different layers to task-specific information preservation.

\section{Background}
\label{sec:background}
\subsection{Learning-based image denoising}
\label{ssec:denoising-network}
End-to-end learning-based denoising methods hold significant potential for medical imaging applications~\cite{zhang2017beyond,yang2018low,li2014adaptive,gong2018pet,chen2017low,jifara2019medical}. 
Given a noisy image $\mathbf{f_n}\in\mathbb{R}^{N}$, where $N$ is the dimension of image, an end-to-end learning-based denoising method can be described generically as: 
\begin{equation}
\hat{\mathbf{f}}
= \mathcal{F}(\mathbf{f_n}; \mathbf{\Theta}),
%=\mathcal{F}(\mathcal{H}{\mathbf{f}} + \mathbf{n}),
\label{eq:denoising}
\end{equation}
where $\mathcal{F}$ denotes an image-to-image mapping implemented by a DNN that is parameterized by the weight vector $\mathbf{\Theta}$
and $\hat{\mathbf{f}}\in\mathbb{R}^{N}$ denotes the denoised image. 
Depending on how the target data are defined when training the DNN, $\hat{\mathbf{f}}$ can be interpreted as an estimate of the noiseless image $\mathbf{f}\in\mathbb{R}^{N}$ or an estimate of {a reduced noise version of $\mathbf{f_n}$}.
A variety of DNNs have been employed to implement the mapping  $\mathcal{F}$ \cite{gong2018pet,jifara2019medical,he2016deep}, and  convolutional neural networks (CNNs) represent a popular choice \cite{manduca2009projection,li2014adaptive,le2013denoising,zhang2017beyond,gong2018pet,you2019denoising}. 
%Networks consisting of multiple convolutional layers have been employed for both natural image \cite{jain2008natural,zhang2018ffdnet} and medical image denoising \cite{gong2018pet, lin2001improving}. 
%In addition to CNN-based methods, a variety of other learning-based approaches have also been developed for medical image denoising by the use of other networks~, 
%e.g., residual neural networks (ResNet)~\cite{he2016deep}. 
%The ResNet employs shortcut connections (the so-called skip connections) 
%between non-adjacent convolutional layers. 
%This network design can better address the vanishing gradient issue \cite{he2016deep}, 
%and allows for a deeper network with more convolutional layers. 

In addition to the choice of DNN architecture, the specification of the loss function plays a key role
in the design of a DNN-based denoising method. 
Mean square error (MSE) that measures the $L^2$ distance between the denoised and target images has been widely employed~\cite{zhang2017beyond,yang2018low,manduca2009projection,li2014adaptive,le2013denoising,chen2017low,manjon2012new,jifara2019medical}. %In addition to the MSE loss, 
The perceptual loss function has also been used and was reported to be effective in reducing noise while retaining image details~\cite{gong2018pet}, and use of {adversarial loss functions have been deployed with similar success}~\cite{wolterink2017generative,yang2018low}. 
However, such loss functions that are commonly employed in computer vision applications do not explicitly incorporate information regarding a particular medical imaging task. 
In recent studies, it has been demonstrated that learning-based denoising methods trained by use of such loss functions can improve traditional IQ measures such as RMSE or SSIM while important information relevant to a downstream detection task is lost \cite{li2021assessing,yu2020ai}.  Such findings motivate the further development
and investigation of task-informed learning-based denoising methods.
%when training DNN-based denoising methods by use of these physical-based loss functions,
%it has been demonstrated that the visibility of important structural details can be compromised or reduced \cite{yang2018low} in the denoised images, despite the fact that traditional image quality metrics (such as RMSE or SSIM) are improved }. As a result, task-relevant information may not be preserved.

\subsection{Formulation of binary signal detection task}
\label{ssec:binary-task}

%In this study, the impact of a task-informed training method on the binary signal detection task performance is assessed.

In this study, a binary signal detection task is considered that requires an observer 
to classify a denoised image $\hat{\mathbf{f}}$ as satisfying either a signal-present hypothesis $H_1$ or a signal-absent hypothesis $H_0$.
%Here, $\hat{\mathbf{f}}$ represents either the denoised image $\mathbf{f}_{denoised}$ or noisy image $\mathbf{f}_{noisy}$.
These two hypotheses can be described as:
\begin{subequations}
	\label{eq:hypo}
	\begin{equation}
	H_0:\hat{\mathbf{f}}=\mathcal{F}(\mathbf{f_{b}+n)} 
%	= \mathcal{H}(\mathcal{O}(\mathbf{f}_{b})),
	\end{equation}
	\begin{equation}
%	H_1:\hat{\mathbf{f}}=\mathbf{f_b+f_s}+\mathbf{n},   
    H_1:\hat{\mathbf{f}}=\mathcal{F}(\mathbf{f_{b+s}+n}),
 %   = \mathcal{H}(\mathcal{O}\mathbf{f}_{(b+s)}),
	\end{equation}   
\end{subequations}
\noindent where $\mathbf{f_{b+s}}\in\mathbb{R}^{N}$ and $\mathbf{f_{b}}\in\mathbb{R}^{N}$ denote signal-present and signal-absent noiseless images and $\mathbf{n}\in\mathbb{R}^{N}$ denotes the measurement noise. A signal-present image $\mathbf{f_{b+s}}$ is formulated by inserting a signal image $\mathbf{f_{s}}\in\mathbb{R}^{N}$ into {a} background object $\mathbf{f_{b}}$.
% $\mathcal{G}(\mathbf{f_{b+s}})$ and $\mathcal{G}(\mathbf{f_{b}})$ denote noisy images with or without signal, $\mathbf{f_{s}}\in\mathbb{R}^{N}$ and $\mathbf{f_{b}}\in\mathbb{R}^{N}$ denote signal and background image, and $\mathcal{G}(\cdot)$ is the mapping between noiseless target image and noisy image.
In a SKE detection task, $\mathbf{f_{s}}$ is non-random, whereas in a SKS detection task it is a random process.
Similarly, in a BKE detection task, $\mathbf{f_{b}}$ is nonrandom, whereas in a BKS detection task it is a random process.
% $\mathcal{F}$ describes the mapping from noisy image to denoised image, 
% and $\hat{\mathbf{f}}$ represents the denoised image.
To perform this task, a deterministic observer computes a test statistic that maps the image 
$\hat{\mathbf{f}}$ to a real-valued scalar test statistic, which is used to perform the task. 
% that is compared to a predetermined threshold $\tau$ to determine if $\hat{\mathbf{f}}$ satisfies $H_0$ or $H_1$. 
% By varying the threshold $\tau$, a ROC curve can be formed to quantify the trade-off between the false-positive fraction (FPF) and the true-positive fraction (TPF) \cite{barrett2013foundations}. The area under the ROC curve (AUC) can be subsequently calculated as a figure-of-merit (FOM) for signal detection performance.

\subsection{Numerical observers for objective IQ assessment}
\label{ssec:observers}

In preliminary assessments of medical imaging technologies, numerical observers (NOs) have been employed to quantify task-based measures of IQ for various image-based inferences\cite{barrett1993model}.
The NOs that are employed in this study are surveyed below.

\subsubsection{Hotelling Observer} 

The Hotelling Observer (HO) employs the Hotelling discriminant, which is the population equivalent of the Fisher linear discriminant, and is optimal among all linear observers in
the sense that it maximizes the signal-to-noise ratio of the test statistic~\cite{barrett2013foundations,fisher1936use}.
The HO test statistic $t_{\text{HO}}(\hat{\mathbf{f}})$ computed by use of the denoised data $\hat{\mathbf{f}}$ is defined as:
\begin{equation}
\label{eq:HO}
t_{\text{HO}}(\hat{\mathbf{f}})=\mathbf{w}^{T}_{\text{HO}}\hat{\mathbf{f}}
=(\mathbf{K_{\hat{f}}}^{-1}\Delta\Bar{\mathbf{f}})^T\hat{\mathbf{f}},
%=(\mathbf{K_g}^{-1}\Delta\Bar{\mathbf{g}})^T\mathbf{g},
\end{equation}
\noindent where $\mathbf{w}^{T}_{\text{HO}}\in\mathbb{R}^{N}$ denotes the Hotelling template,
$\Delta\Bar{\mathbf{f}}\in\mathbb{R}^{N}$ denotes the difference between the ensemble mean of 
the image data $\hat{\mathbf{f}}$ under the two hypotheses $H_0$ and $H_1$, and
$\mathbf{K}_{\hat{\mathbf{f}}} \equiv \frac{1}{2}(\mathbf{K}_{0}(\hat{\mathbf{f}})+\mathbf{K}_{1}(\hat{\mathbf{f}}))$.
Here, $\mathbf{K}_{0}(\hat{\mathbf{f}}) \in \mathbb{R}^{{N}\times {N}}$ and $\mathbf{K}_{1}(\hat{\mathbf{f}})\in \mathbb{R}^{{N}\times {N}}$ 
denote the covariance matrices corresponding to $\hat{\mathbf{f}}$ under $H_0$ and $H_1$, respectively.

In some cases, the covariance matrices 
$\mathbf{K}_{0}(\hat{\mathbf{f}})$ and $\mathbf{K}_{1}(\hat{\mathbf{f}})$ are ill-conditioned and therefore their inverse cannot be stably computed.
To address this, a regularized HO (RHO) can be employed that implements the test statistic $t_{\text{RHO}}(\hat{\mathbf{f}})$ as \cite{li2021assessing}:
\begin{equation}
	\label{eq:RHO}
	t_{\text{RHO}}(\hat{\mathbf{f}})= \mathbf{w}^{T}_{\text{RHO}}\hat{\mathbf{f}}  =(\mathbf{K}_{\alpha}^{+}	\Delta\Bar{\mathbf{f}})^T\hat{\mathbf{f}},
\end{equation}
where $\mathbf{K}_{\alpha}$ represents a low-rank approximation of $\mathbf{K}_{\hat{\mathbf{f}}}$ that is formed by keeping only the singular values  of $\mathbf{K}_{\hat{\mathbf{f}}}$ greater than  
$\alpha\sigma_{max}$.  
Here, 
$\alpha$ is a tunable parameter  and $\sigma_{max}$ represents the largest singular value of $\mathbf{K}_{\hat{\mathbf{f}}}$.
 %The value of $\alpha$ can be tuned on an independent set of data and the value that leads to the best RHO performance can be selected.
 Finally, $\mathbf{K}_{\alpha}^{+}$ is the Moore–Penrose inverse of $\mathbf{K}_{\alpha}$.

It has been demonstrated that when data with large dimensions are considered, the estimation and inversion of the covariance matrix can be intractable \cite{barrett2001megalopinakophobia}, which makes the implementation of the HO challenging. To circumvent this difficulty, a supervised learning-based method that uses a simple single layer neural network  (SLNN) to approximate the HO has been proposed recently \cite{zhou2019approximating}. The HO discriminant function can be modeled by a SLNN that possesses only a single fully-connected layer. This SLNN-approximated HO (SLNN-HO) directly learns the Hotelling template without explicitly estimating and inverting covariance matrices. It has been reported that the SLNN-HO is effective with large images and a limited number of images \cite{zhou2019approximating}. 

%\vspace{0.1cm}
\subsubsection{Channelized Hotelling observer}
A channelized HO (CHO) is formed when the HO is employed with a channeling mechanism.
When implemented with anthropomorphic channels and an internal noise mechanism, the CHO can be interpreted as an anthropomorphic observer and attempts to predict human observer performance~\cite{myers1987addition,abbey2001human}.
In addition, the channeling mechanism can also be employed to reduce the dimensionality of the image data when the image data is insufficient to accurately estimate the covariance matrix.
% A channelized HO (CHO) is formed when the HO is employed with a channeling mechanism to reduce the dimensionality of the image data.
% The CHO with difference-of-Gaussian (DOG) channels and an internal noise mechansim can be interpreted as an anthropomorphic observer~\cite{myers1987addition,abbey2000modeling,abbey2001human}.
% When the imaging data is insufficient to accurately estimate the covariance matrix or an anthropomorphic CHO is needed, the channeling mechanism can be employed. 
Let $\mathbf{T}$ denote a channel matrix and $\mathbf{v\equiv T}\hat{\mathbf{f}}$ the corresponding channelized image data. The CHO test statistic $t_{\text{CHO}}(\hat{\mathbf{f}})$ is given by:
\begin{equation}
\label{eq:CHO}
t_{\text{CHO}}(\hat{\mathbf{f}})=\left[(\mathbf{K_v+K_{int}})^{-1}\Delta\Bar{\mathbf{v}}\right]^T(\mathbf{v+v_{int}}),
\end{equation}
where $\mathbf{K_v}$ denotes the covariance matrix of the channelized data $\mathbf{v}$, 
$\mathbf{K_{int}}$ denotes the covariance matrix of the channel internal noise, 
and $\mathbf{v_{int}}$ is a noise vector sampled from {a} Gaussian distribution $\mathcal{N}(\mathbf{0,K_{int}})$. 
Based on previous studies \cite{abbey2001human}, in this work $\mathbf{K_{int}}$ is defined as:
\begin{equation}
\label{eq:int_noise}
\mathbf{K_{int}}=\epsilon \cdot diag(\mathbf{K_v}),
\end{equation}
where $diag(\mathbf{K_v})$ represents a diagonal matrix with diagonal elements from $\mathbf{K_v}$ 
and $\epsilon$ is the internal noise level. The parameters of the difference-of-Gaussian (DOG) channels and the internal noise level 
employed in this study are described below in Section~\ref{ssec:numerical_observers}.

%\vspace{0.1cm}
\subsubsection{Learned NOs} 

% For binary signal detection tasks,
% the Bayesian Ideal Observer (IO) sets an upper limit of observer performance.
% Except in special cases, the determination of the IO test statistic is analytically intractable. 
%Markov-Chain Monte Carlo (MCMC) techniques can be employed to approximate the IO detection performance, but their reported applications have been limited to relatively simple object models.
Recently, several machine learning methods have been proposed to establish NOs \cite{zhou2019approximating,zhou_lroc, li2021hybrid}.
% Zhou et al employed convolutional neural networks (CNNs) \cite{zhou2019approximating} and single-layer neural
% networks (SLNNs) to approximate the IO test statistics.
The SLNN-based NO (SLNN-NO) is a special learned NO that has the shallowest architecture that possesses only a single fully-connected layer with a bias term and a sigmoid activation function. The binary cross entropy (BCE) loss function can be used to train the SLNN-NO. 
% In most cases, 
% Considering the insufficient model capacity of the SLNN-NO in most cases, the SLNN-NO is a sub-optimal observer.

\section{Task-informed training method}
\label{sec:methods}
In this work, a transfer learning approach is
investigated in which {a} DNN is pre-trained by use of a conventional (non-task-informed) loss function $\mathcal{L}_{\text{p}}$ and subsequently fine-tuned by use of a hybrid loss $\mathcal{L}_{\text{Hybrid}}$ that includes a task-component $\mathcal{L}_{\text{t}}$. The fine-tuning of the denoising network is constrained to the last several layers instead of re-training the whole  network.
This  is motivated by a recent study by Li \etal \cite{li2021assessing} that demonstrated, at least for linear CNN-based denoising networks, the degradation of task-relevant information primarily occurs in the last layers. A hybrid loss function $\mathcal{L}_{\text{Hybrid}}$ 
%that is a convex combination of physical and task-based loss components is employed to fine-tune the pretrained network,
is defined as \cite{adler2021task}:
\begin{equation}
\label{eq:hloss}
\mathcal{L}_{\text{Hybrid}}(\mathbf{\Theta}_1,\mathbf{\Theta}_o)=\lambda\cdot\mathcal{L}_{\text{p}}(\mathbf{\Theta}_1)+(1-\lambda)\cdot\mathcal{L}_{\text{t}}(\mathbf{\Theta}_1,\mathbf{\Theta}_o),
\end{equation}
where $\lambda\in[0,1]$ is a scalar parameter,  $\mathcal{L}_{\text{p}}$ is the physical loss component, $\mathcal{L}_{\text{t}}$ is the task-component, $\mathbf{\Theta}_1$ is the vector of weight parameters associated with the trainable layers in the pretrained denoising network, and $\mathbf{\Theta}_o$ denotes the vector of weight parameters of the NN-based NO used to compute the task-component $\mathcal{L}_{\text{t}}$.
The task-component is designed to measure the performance of a NO on a specific task. By appending a {network-based} NO to the pretrained denoising network, the denoised image can be transformed into a scalar that is used to compute the task-specific component.
The {trainable} layers in the pretrained denoising network are jointly trained with the NO. By employing this training strategy, the NO used to compute $\mathcal{L}_{\text{t}}$ can be easily adapted to different tasks.
The details of the proposed task-informed training method are described below and summarized in Algorithm~\ref{alg:method}.

Mean-squared-error (MSE) and mean-absolute-error (MAE) are commonly employed choices for $\mathcal{L}_{\text{p}}$. The selection of the task-based loss component is based on specific tasks. In this paper, binary signal detection tasks were considered, {and} the specific formulation of $\mathcal{L}_{\text{Hybrid}}$ is described in Section \ref{ssec:DNN_intro}.

\begin{algorithm}[ht]
	\DontPrintSemicolon
	\KwInput{
	\begin{itemize} 
	\item{$\mathbf{\Theta}$: Weight parameters of the denoising network $\mathcal{F}$ with $N$ layers;}
	
  \item{$\mathbf{\Theta}_1$: Weight parameters of the last $N_{train}$ trainable layers of $\mathcal{F}$;}  
    \item {$\mathbf{\Theta}_2$: Weight parameters of the first $(N-N_{train})$ layers of the denoising network $\mathcal{F}$, $\mathbf{\Theta}=\{\mathbf{\Theta}_1, \mathbf{\Theta}_2$\}; }
  \item{$\mathbf{\Theta}_o$: Weight parameters of an appended observer used to compute the task-component $\mathcal{L}_{\text{t}}$;}

 % ($\mathbf{\Theta}_1\subseteq\mathbf{\Theta}$);}
    
    \item{$\mathcal{L}_\text{p}$: Physical loss for pre-training and task-informed training $\mathcal{F}$;}
    
    \item{$\mathcal{L}_\text{t}$: Task-based loss for task-informed training $\mathcal{F}$;}
    
    \item{$\mathcal{L}_\text{Hybrid}$: The hybrid loss formulated by $\mathcal{L}_\text{p}$ and $\mathcal{L}_\text{t}$, and weighted by the parameter $\lambda$ defined in Eqn. (\ref{eq:hloss});}
    
    \item{$\mathcal{D}_1$: Dataset for pre-training $\mathcal{F}$;}
    
    \item{$\mathcal{D}_2$: Dataset for task-informed training $\mathcal{F}$;}
	\end{itemize} 
	}
	\KwOutput{The denoising network $\mathcal{F}$ with optimized weight parameters $\mathbf{\Theta}$ after task-informed training by use of $\mathcal{L}_\text{Hybrid}$.} 
	\begin{enumerate}
	
	\item {
	Given initial setting of parameters $\mathbf{\Theta}$, pretrain the denoising network $\mathcal{F}$ and optimize weight parameters $\mathbf{\Theta}$ by use of $\mathcal{D}_1$ and physical loss function $\mathcal{L}_{\text{p}}$;}\\
	% \tcp{The network weight parameters $\mathbf{\Theta}$ are not optimized for specific task in this step.}
	
	\item{
	Append the observer with weight parameters $\mathbf{\Theta}_o$ to the pretrained $\mathcal{F}$;}
	
	\item{
	Set the weight parameters $\mathbf{\Theta}_1$ that correspond to the last $N_{train}$ convolutional layers of the pretrained $\mathcal{F}$ to be trainable; %and fix $\mathbf{\Theta}_2$;
	}\\
	 \tcp{The other trained weight parameters $\mathbf{\Theta}_2$ are fixed;}
	 %{, where $\mathbf{\Theta} = \{\mathbf{\Theta}_1, \mathbf{\Theta}_3\}$}
	
	\item{Given initial setting of $\mathbf{\Theta}_o$ and pre-trained $\mathbf{\Theta}$, jointly tune the weight parameters $\mathbf{\Theta}_1$ and train $\mathbf{\Theta}_o$ by use of $\mathcal{D}_2$ and the hybrid loss function $\mathcal{L}_{\text{Hybrid}}$;}\\ 
	% \tcp{Only the weight parameters $\mathbf{\Theta}_1\cup \mathbf{\Theta}_o$ are trainable in this step}
	
	\item{Output the task-informed denoising network $\mathcal{F}$ with optimized weight parameters $\mathbf{\Theta}$.}\\
	%\tcp{The weight parameters $\mathbf{\hat{\Theta}}$ is a task-informed version of weight parameters $\mathbf{\Theta}$ in step $\mathbf{1}$}
	
	\end{enumerate}

	\caption{General procedure of the task-informed training method}
 
	\label{alg:method}

\end{algorithm}

%To fine-tune the denoising network using the hybrid loss strategy, a classifier is appended to the pretrained denoising network. In this way, the denoised image can be transformed into a scalar that is used to compute the task-specific component. In addition, only the last several convolutional layers in the pretrained denoising network are set to be trainable.

\section{Numerical studies}
\label{sec:numerical_studies}

In this paper, the task-informed training method {in Algorithm~\ref{alg:method}} was objectively assessed in a stylized low-dose X-ray computed tomography (CT) denoising study. 
Binary signal detection tasks under signal-known-statistically (SKS) with background-known-statistically (BKS) conditions were considered. 
Both the SLNN-NO and SLNN-HO were considered as the NOs {employed} to compute the task-component $\mathcal{L}_\text{t}$ in Eqn. (\ref{eq:hloss}). 
The performance of {the SLNN-NO, SLNN-HO, and other NOs described in Sec. \ref{ssec:observers}} on denoised images was quantified to evaluate the impact of task-informed training procedure on the denoising network. 
The details of numerical studies are described below.

\subsection{Virtual imaging pipeline}
\label{ssec:data_generation}

The Lung Image Database Consortium image collection
(LIDC) \cite{armato2011lung} was employed to generate signal-present (SP) images $\mathbf{f_{b+s}}$ and signal-absent (SA) images $\mathbf{f_{b}}$ to perform binary signal detection tasks {as defined} in Eqn. (\ref{eq:hypo}).
This database consists of 243,945 2D image slices from 1,018 3D thoracic CT reconstructed images, 
in which 10,706 image slices contain annotated nodules. In addition to {the} 10,706 {SP images}, additional {SP images} were generated by inserting realistic nodules using an established insertion method
\cite{pezeshk2014seamless}. 
%and the lesions are marked up. 
% The SA and SP images used for training and evaluation were generated as follows.
% For SA images, 
Fifty-thousand {SA images} were formed by extracting regions-of-interest (ROIs) of dimension of $120\times120$ pixels from normal lung areas from central several slices. 
%Similarly, ROIs of the same dimension that contain lesions (nodules) were extracted and considered as SP images.
%In these SP images, the centroids of the nodules were either fixed or random subject to specific tasks described in Sec. \ref{ssec:evaluation_detail}.
The same number of {SP images} were generated.
%is employed to insert realistic nodules into signal-absent images.
%additional SP images were formed by inserting nodules into \textcolor{blue}{SA images, maybe we should say images without nodules?} by use of 
In {SP images}, the centroids of the nodules were either located at a fixed location or at random locations subject to the specific tasks described in Section \ref{ssec:evaluation_detail}.
%\textcolor{blue}{The centroids of the inserted lesions were the same as described above. 
% The calculated mean values of the original and simulated SP images were 0.2425 and 0.2156, respectively, which indicates the statistical similarity of original and simulated SP images.
%to validate the similarity. 
%that The statistic of simulated SP images was computed and compared with that of original SP images to validate the similarity. the image difference is within a reasonable range.
The generated {SP and SA} images were utilized as the target (normal-dose) CT images $\mathbf{f}$. 

\begin{figure}[h]
     \centering
     \includegraphics[width=0.42\textwidth]{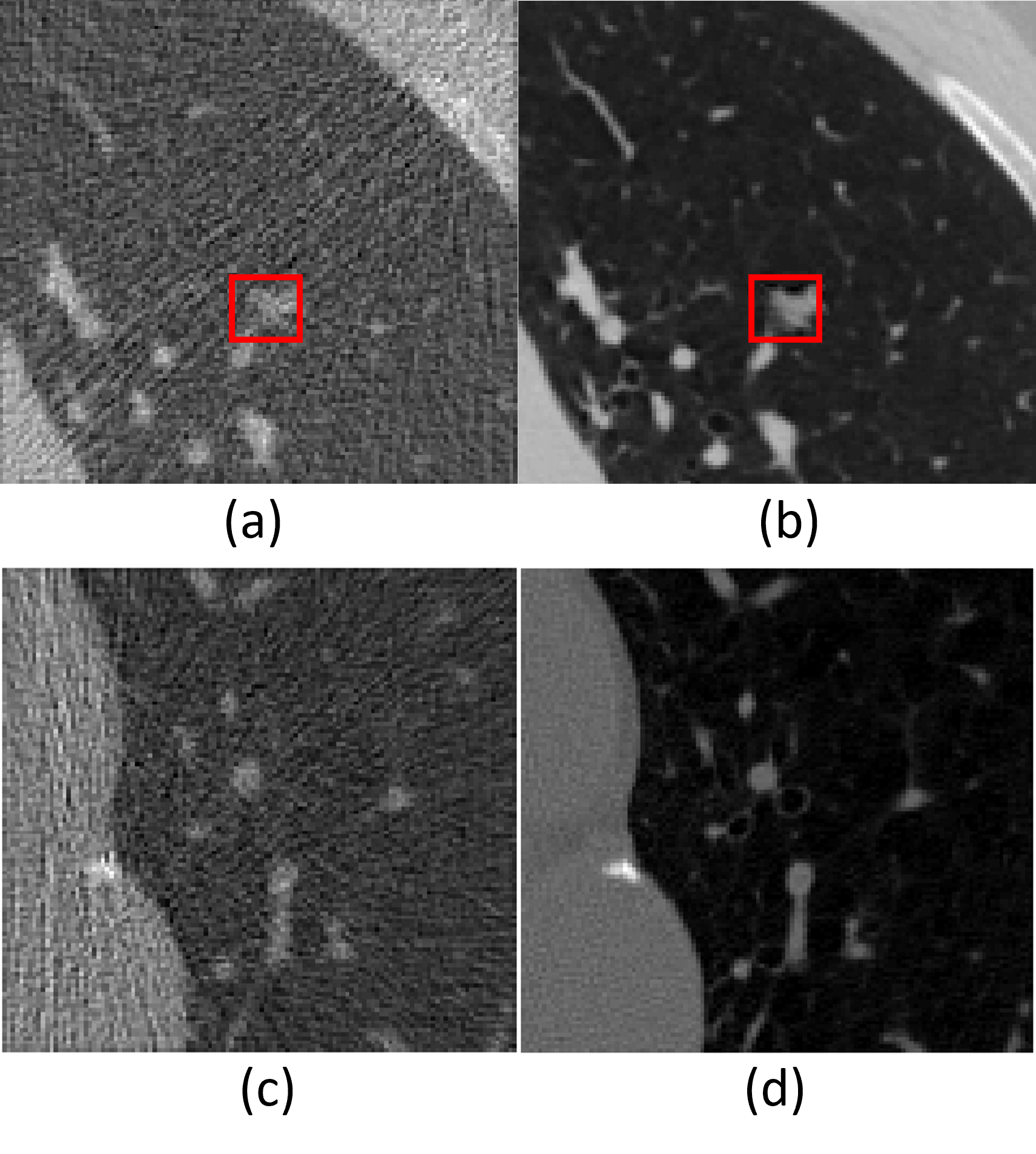}
    \caption{Examples of (a) noisy (low-dose) signal-present image, (b) target (normal-dose) signal-present image, (c) noisy (low-dose) signal-absent image, and (d) normal-dose signal-absent image. The red box contains the signal.
    }
    \vspace{-0.1in}
    \label{fig:sample}
\end{figure}

The corresponding noise-enhanced (low-dose) images $\mathbf{f_{n}}$ were generated {by degrading the} target images $\mathbf{f}$ described above.
%for training the denoising method, 
The target image $\mathbf{f}$ was first transformed to the projection domain by using the 2D discrete Radon transform $\mathcal{R}$. 
% \textcolor{blue}{[Delete] $\{$The sinogram was transformed into transmission data \cite{zeng2015simple} and was multiplied by the incident flux $I_0$ that controls the noise level in the generated low-dose image. Poisson noise was subsequently added in the projection domain to generate noise-enhanced projections.$\}$}
Noise-enhanced projection $\mathbf{\hat{g}}$ were generated by adding Poisson noise to the transmission data, which is described as:
	\begin{equation}
    \mathbf{\hat{g}}=\mathcal{T}^{-1}(Poi(\mathcal{T}(\mathcal{R}\mathbf{f}))).  
      	\label{eq:add_noise}
    \end{equation}
Here, $Poi(\cdot)$ was a Poisson random vector, $\mathcal{T}(\mathbf{x}) = I_0\exp(-\frac{\mathbf{x}}{n})$, and $\mathcal{T}^{-1}(\mathbf{x}) = -n\log(\frac{\mathbf{x}}{I_0})$, 
% controlled by the parameter $I_0$.  
where $n$ is a scalar used for normalization and the parameter $I_0$ controlled noise level \cite{zeng2015simple}.
The noisy (low-dose) images $\mathbf{f_{n}}$ were then reconstructed by use of a filtered back-projection (FBP) reconstruction algorithm that employed a Ram-Lak filter \cite{kak2001principles}.
Samples of the generated images are shown in Fig. \ref{fig:sample}.
The pairs of noisy and target images are used for training the denoising method described below in Section \ref{ssec:DNN_intro}. 

%\subsection{DNN-based denoising methods, training, and validation}
\subsection{Training and validation details}
\label{ssec:DNN_intro}
\subsubsection{Architecture and loss function for denoising networks}
\begin{figure}[h]
     \centering
     \includegraphics[width=0.5\textwidth]{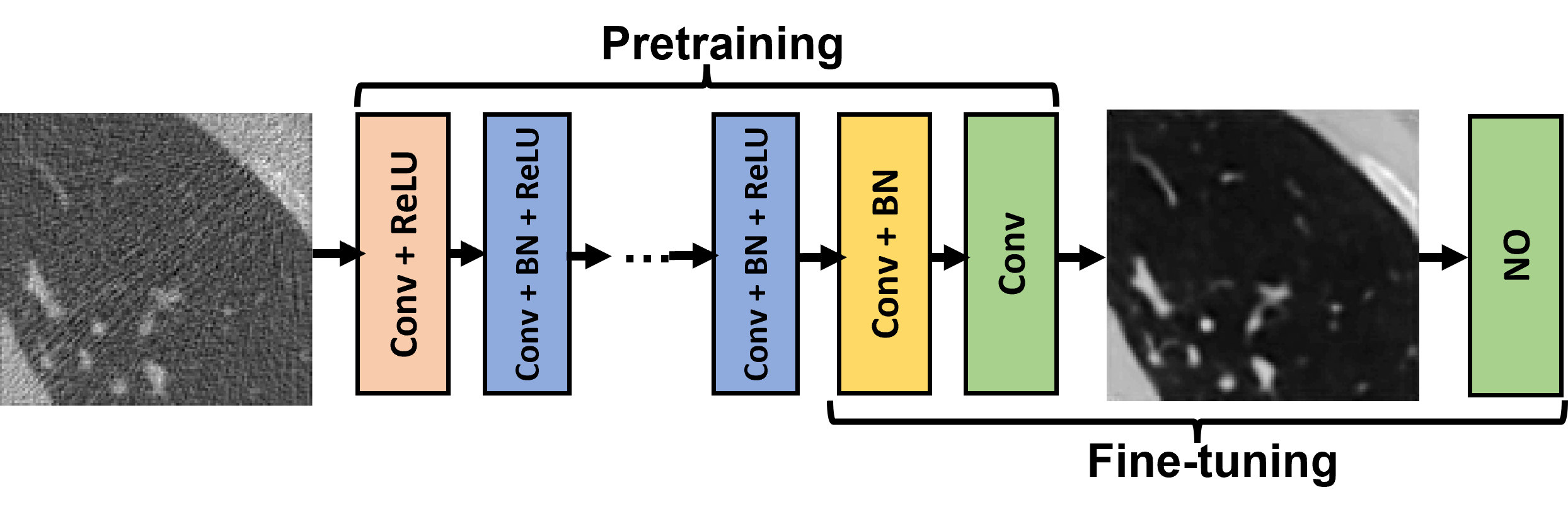}
    \caption{A CNN-based denoising network was investigated in this study.}
    % \vspace{-0.1in}
    \label{fig:CNN_architecture}
\end{figure}

The canonical CNN architecture of depth $D$ depicted in Fig.~\ref{fig:CNN_architecture} was employed with the task-informed training method to establish an end-to-end learned denoising method. 
It is important to note that the  assessment studies described below can be readily repeated with any other DNN.
%As depicted in Fig.~\ref{fig:CNN_architecture}, a CNN architecture of depth $D$ was considered.
The network input was a reconstructed noisy image $\mathbf{f}_{n}$ of dimension $120\times 120$
and the output was a denoised image $\hat{\mathbf{f}}$ with the same dimensions.
The CNN contained four types of layers.
The first layer was a Conv+ReLU layer, in which 64 convolution filters of dimension $3\times3\times1$ were applied to generate 64 feature maps.
In each of the $2^{nd}$ to $(D-2)^{th}$ Conv+BN+ReLU layers, 
64 convolution filters of dimension $3\times3\times64$ were employed 
and batch normalization was included between the convolution and ReLU operations. 
In the $(D-1)^{th}$ Conv+BN layer, 64 convolution filters of dimension $3\times3\times64$ were employed and batch normalization was performed. 
In the last Conv layer, one single convolution filter of dimension $3\times3\times64$ was employed to form the final denoised image of dimension $120 \times 120$.

Let $\mathbf{f}^{(j)}$ denote a given SA or SP target (normal-dose) image and let $\mathbf{f_{n}}^{(j)}$ denote the corresponding noise-enhanced (low-dose) image.
Given a collection of $J$ paired training data
$\{(\mathbf{f}_{n}^{(j)},\mathbf{f}^{(j)})\}^J_{j=1}$, 
the denoising network was pretrained by minimizing the MSE loss function: %$\mathcal{L}_{\text{MSE}}(\mathbf{\Theta})=\frac{1}{J}\sum_{j=1}^J \Vert \mathcal{F}(\mathbf{f}_{noisy}^{(j)};	\mathbf{\Theta})-\mathbf{f}^{(j)}\Vert_2^2.$
\begin{equation}
\label{eq:MSE}
\mathcal{L}_{\text{MSE}}(\mathbf{\Theta})=\frac{1}{J}\sum_{j=1}^J \Vert \mathcal{F}(\mathbf{f}_{n}^{(j)};	\mathbf{\Theta})-\mathbf{f}^{(j)}\Vert_2^2,
\end{equation}
where the vector $\mathbf{\Theta}$ denotes the weight parameters of the denoising network.

\subsubsection{Architecture and loss function for the NN-based observers used to compute $\mathcal{L}_\text{t}$}
\label{sssec:appended_NO}
The physical loss function $\mathcal{L}_{\text{p}}$ in Eqn. (\ref{eq:hloss}) was defined by a MSE loss.
The task-component $\mathcal{L}_{\text{t}}$ of the hybrid loss function was computed by use of either the SLNN-NO or SLNN-HO as described next.
%These network-based NOs were appended to the pretrained denoising network and the hybrid loss function $\mathcal{L}_{\text{Hybrid}}$ in Eqn. (\ref{eq:hloss}) was used to jointly fine-tune the pretrained denoising network and train the observer. 

The SLNN-NO consisted of a fully-connected layer along with a sigmoid activation function. The BCE loss function was employed to train the SLNN-NO. Let $\{(\mathbf{f}_{n}^{(j)},y^{(j)})\}^J_{j=1}$ denote the image data $\mathbf{f}_{n}^{(j)}$ and the corresponding label $y_j\in\{0,1\}$. The BCE loss function $\mathcal{L}_{\text{BCE}}(\mathbf{\Theta}_1,\mathbf{\Theta}_o)$ can be expressed as \cite{zhou2019approximating}:
\begin{equation}
\label{eq:bce}
\mathcal{L}_{\text{BCE}}(\mathbf{\Theta}_1,\mathbf{\Theta}_o)=-\sum_{j=1}^{J} \log p(y_j|\mathbf{f}_{n}^{(j)},\mathbf{\Theta}_1,\mathbf{\Theta}_o).
\end{equation}
Here, $\mathbf{\Theta}_1$ is the vector of weight parameters associated with the trainable layers in the pretrained denoising network, and the vector $\mathbf{\Theta}_o$ denotes the weight parameters of the fully-connected layer of the appended SLNN-NO.

% As described in Section \ref{ssec:observers}, the HO can be approximated by a SLNN that only possessed a single fully connected layer. The corresponding HO loss function was employed. 
% The vector of weight parameters in the SLNN was denoted as $\mathbf{\Theta}_o\in\mathbb{R}^{N\times1}$.
% By using the notations defined above, 
Differently. the SLNN-HO loss function $\mathcal{L}_{\text{HO}}(\mathbf{\Theta}_1,\mathbf{\Theta}_o)$ can be expressed as \cite{zhou2019approximating}:
% \begin{equation}
\begin{align}
\label{eq:HO_loss}
\mathcal{L}_{\text{HO}}(\mathbf{\Theta}_1,\mathbf{\Theta}_o)= \,& \frac{1}{J}\sum_{j=1}^J \Bigl\{ (1-y_j)[ \mathbf{\Theta}_o^T(\mathcal{F}(\mathbf{f}_{n}^{(j)};	\mathbf{\Theta}_1) - \mathbf{\Bar{f}}_0) ]^2\notag \\ 
&+ y_j[ \mathbf{\Theta}_o^T(\mathcal{F}(\mathbf{f}_{n}^{(j)};	\mathbf{\Theta}_1) - \mathbf{\Bar{f}}_1) ]^2 \Bigr\}  \\ 
&-2\mathbf{\Theta}_o^T\Delta \mathbf{\Bar{f}},\notag
\end{align}
% \raisetag{-10pt}
% \end{equation}
where $\mathbf{\Bar{f}}_0 = \frac{2}{J}\sum_{j=1}^J(1-y_j)\mathcal{F}(\mathbf{f}_{n}^{(j)};	\mathbf{\Theta}_1)$, $\mathbf{\Bar{f}}_1 = \frac{2}{J}\sum_{j=1}^J y_j\mathcal{F}(\mathbf{f}_{n}^{(j)};	\mathbf{\Theta}_1)$, and $\Delta \mathbf{\Bar{f}} = \mathbf{\Bar{f}}_1 - \mathbf{\Bar{f}}_0$.

% Considering a joint detection and localization (detection-localization) task, a softmax activation function and the categorical cross entropy (CCE) loss function can be employed to formulate the SLNN-NO. Using the same notations described above, the CCE loss function $\mathcal{L}_{\text{CCE}}(\mathbf{\Theta}_1,\mathbf{\Theta}_o)$ can be expressed as:
% \begin{equation}
% \label{eq:cce}
% \mathcal{L}_{\text{CCE}}(\mathbf{\Theta}_1,\mathbf{\Theta}_o)=-\frac{1}{J}\sum_{j=1}^{J} \log p(H_{y_j}|\mathbf{f}_{n}^{(j)},&\mathbf{\Theta}_1,\mathbf{\Theta}_o).
% \end{equation}

\subsubsection{Datasets and denoising network training details}
% To pretrain the denoising network, the MSE loss function in Eqn. (\ref{eq:MSE}) was employed. 
The standard convention of utilizing separate training/validation/testing datasets was adopted. The training dataset 
included 40,000 pairs of noisy signal-present and signal-absent images along with the corresponding target (normal noise) images. The validation dataset including 200 signal-present images and 200 signal-absent images and the corresponding target (normal noise) images was randomly selected from the training dataset. 
Finally, the testing dataset comprised 10,000 signal-present images and 10,000 signal-absent noisy images.
For task-informed model training with the hybrid loss function $\mathcal{L}_{Hybrid}$, 
%the observer was appended to the pretrained denoising network and the hybrid loss function in Eqn. (\ref{eq:hloss}) was used.
the same training dataset described above was employed to fine-tune the denoising network.
%to fully optimize the trainable layers in the pretrained model with the parameters of other layers fixed. 
The validation and testing datasets used for pretraining were also employed to evaluate the performance of the fine-tuned denoising networks.

In both the pretraining and task-informed fine-tuning stages, the denoising networks were trained on mini-batches at each iteration by use of the Adam optimizer~\cite{kingma2014adam} with a learning rate of 0.0001.
Each mini-batch contained 50 signal-present images and 50 signal-absent images that were randomly selected from the training dataset.
The network model that possessed the best performance on the validation dataset was selected for use. 
The Keras library~\cite{chollet2015keras} was employed for implementing and training all networks on a single NVIDIA TITAN X GPU.

\subsection{Objective evaluation of image quality}
\label{ssec:evaluation_detail}
\subsubsection{SKS/BKS binary signal detection tasks with fixed signal locations}
\label{sssec:l_known}

The task-informed training method was evaluated for SKS/BKS binary signal detection tasks where known signal locations were considered. 
% In the detection task, the signal was known statistically as each inserted signal was randomly selected from 13,706 signal realizations. 
The centroids of the nodules were located at the center of the extracted ROIs. The incident flux $I_0=500$ was used to determine the noise level in simulated noisy images. 

Both the SLNN-NO and SLNN-HO were employed as NOs used to compute task component $\mathcal{L}_{\text{t}}$ in Eqn. (\ref{eq:hloss}). 
The SLNN-NO, HO, RHO, and DOG-CHO were employed for subsequent assessments of image quality. 
It should be noted that for the case where the SLNN-NO was employed to compute $\mathcal{L}_{\text{t}}$, the SLNN-NO employed for objective image quality assessment was trained on the {denoised} estimates and it was not identical to that used to compute $\mathcal{L}_{\text{t}}$.
% that was weighted by $\lambda$. 
% This case represented a situation that the different observers were used for training and evaluation.
The weight parameters $\lambda=\{0.01, 0.1, 0.3, 0.5, 0.7, 0.9, 0.99\}$ in Eqn. (\ref{eq:hloss}) were considered.
Only the last three convolutional layers of the denoising network were set to be trainable for both cases.
Based on these settings, the impact of the weight parameter $\lambda$ on the performance of considered NOs was investigated.

%\textcolor{blue}{The observers used for training and evaluation were also mismatched to simulate a situation where the ultimate reader is unknown. (this is for task shift?)}

%\textcolor{blue}{maybe add a couple of sentences to explain the details of mismatch, which mismath  }

\subsubsection{SKS/BKS binary signal detection tasks with random signal locations}
\label{sssec:l_random}
%The task-informed method was evaluated when considering SKS/BKS binary signal detection tasks with random signal locations. 
%In the detection task, 
In this case, the centroids of the nodules were randomly located within the lung area of extracted ROIs by use of a uniform probability density function. The incident flux $I_0=500$ was used to determine the noise level of the simulated low-dose images. 
The SLNN-NO was used to compute task component $\mathcal{L}_{\text{t}}$ in Eqn. (\ref{eq:hloss}), considering that the SLNN-NO can be employed when signal is randomly located.
The trained SLNN-NO was subsequently utilized to evaluate the performance of fine-tuned denoising networks. This represented a situation that the same observer was used for both training and evaluation.

%\textcolor{blue}{((It should be noted that the SLNN-NO for evaluation was independently trained so it is different from the SLNN-NO used for training).(also, how about SLNN-HO?), 
% (also, the observer was retrained?)}% that was weighted by $\lambda$. 

To assess the impact of the weight parameter $\lambda$ on the performance of the SLNN-NO, the weight parameters $\lambda=\{0.01, 0.1, 0.3, 0.5, 0.7, 0.9, 0.99\}$ in Eqn. (\ref{eq:hloss}) were considered. {The number of trainable layers was also swept from 0 to 4.}

% This case mainly focused on the evaluation of several parameters of the task-informed training method to gain insights into their potential impact on task performance.\textcolor{blue}{(I moved this paragraph from the above to here.)} 
% These \textcolor{blue}{experimental settings, (remove this sentence?) were designed to investigate the case where the observers for training and evaluation were matched.}

\subsubsection{Investigation of the task shifts in considered binary signal detection tasks}

%These \textcolor{blue}{experimental settings} were designed to investigate the case where the observers for training and evaluation were matched. 
%\textcolor{blue}{The observers used for training and evaluation were also mismatched to simulate a situation where the ultimate reader is unknown. (this is for task shift?)}

A study was designed to investigate the robustness of {the} task-informed image denoising method to task-shift.
%In clinic, the exact task might not be known.
% This study was designed to simulate cases in which the training dataset cannot accurately describe the clinical situations, i.e., the exact task is unknown. 
% Binary signal detection tasks with fixed and random signal locations were considered to investigate the impact of task-shift in two different situations.
%specific cases were considered. 
%For the first case, the simple task was used for training and the complex one was used for evaluation. 
First, binary signal detection tasks with fixed signal locations were considered for model training (source tasks) while tasks with random signal locations were considered for evaluation (target tasks). 
Next, 
the tasks with random signal locations were used as source tasks and the tasks with fixed locations were considered as target tasks.
%For the second case, the tasks with random signal locations were employed for training and the tasks with fixed signal locations were employed for evaluation.
The SLNN-NO was employed to compute task component $\mathcal{L}_{\text{t}}$ in Eqn. (\ref{eq:hloss}) and only the last three convolutional layers were set to be trainable. 
For evaluations, SLNN-NOs were independently trained on training datasets for target tasks. 
The SLNN-NO performance under the situations without task shift was considered as the reference.
The weight parameters $\lambda=\{0.01, 0.1, 0.3, 0.5, 0.7, 0.9, 0.99\}$ in Eqn. (\ref{eq:hloss}) were considered to investigate the impact of task shifts when the weight of task-based component varies.
% Only the last three convolutional layers were set to be trainable in this case. 

% \textcolor{blue}{Kaiyan added more content, and put in the supplemental document} Detection-localization tasks with 2 and 4 possible signal locations were also considered for investigation. 

\subsubsection{Numerical observer computation}
\label{ssec:numerical_observers}
% \textcolor{blue}{somehow, here the observer for training and observer for evaluation are confused. I tried to clean up. double check and let's discuss.}

% For the first case described above in Section \ref{sssec:l_known}, the SLNN-NO for evaluation was independently trained and 40,000 signal-present images and 40,000 signal-absent images were employed.
% For the second case described in Section \ref{sssec:l_random} above,
%for (\textcolor{blue}{SKS/BKS binary signal detection tasks with random signal locations, I know it is redundant, but think it is more clear}), 
% the SLNN-NO was trained together with the fine-tuning of the denoising network. The SLNN-NO was subsequently utilized to evaluate the performance of fine-tuned denoising networks. 

Both {the} HO and RHO were employed {for objective image quality assessment} since they are optimal linear observers. For computing the HO and RHO test statistics, the covariance matrix $\mathbf{K_{\hat{f}}}$ was empirically estimated by use of 40,000 signal-present and 40,000 signal-absent images. When computing the RHO test statistic, the threshold parameter $\alpha$ in Eqn.~(\ref{eq:RHO}) was swept from $1e-1$ to $1e-7$ and the corresponding detection performance was estimated based on a separate validation dataset including 200 signal-present images and 200 signal-absent images. The value which led to the best RHO detection performance was selected.

For computing the CHO test statistic, 2,000 signal-present and 2,000 signal-absent images were utilized to empirically estimate the channelized covariance matrix. 
A set of 10 DOG channels~\cite{abbey2001human} was employed with channel parameters $\sigma_0=0.005$, $\alpha=1.4$, and $Q=1.67$. 
The internal noise level $\epsilon$ was 2.5, which was the same value employed by Abbey~\etal \cite{abbey2001human}.

To independently train the SLNN-NO {for objective image quality assessment}, 40,000 signal-present images and 40,000 signal-absent images were employed. These learned-NOs were trained by use of the Adam optimizer~\cite{kingma2014adam} with a learning rate of 0.0001.

\subsubsection{Evaluation metrics}

Both traditional and task-based measures of IQ were employed for assessments. 
%To evaluate the performance of the NOs, 
Receiver operating characteristic (ROC) analysis was conducted and
area under the curve (AUC) values were computed and employed as a figure-of-merit for task-based measures.
The ROC curves were fit by use of the Metz-ROC software~\cite{metz} that employs the proper binormal model~\cite{pesce2007reliable}.
The uncertainty of {the AUC} values were estimated as well. Two commonly used traditional metrics (i.e., RMSE and SSIM) were employed as task-agnostic measures to assess the denoised images.

\section{Results}
\label{sec:results}
\subsection{Results for the case with fixed signal locations}
\subsubsection{Impact of the weight parameter $\lambda$}

\begin{figure}[h]
	\centering
	\includegraphics[width=0.48\textwidth]{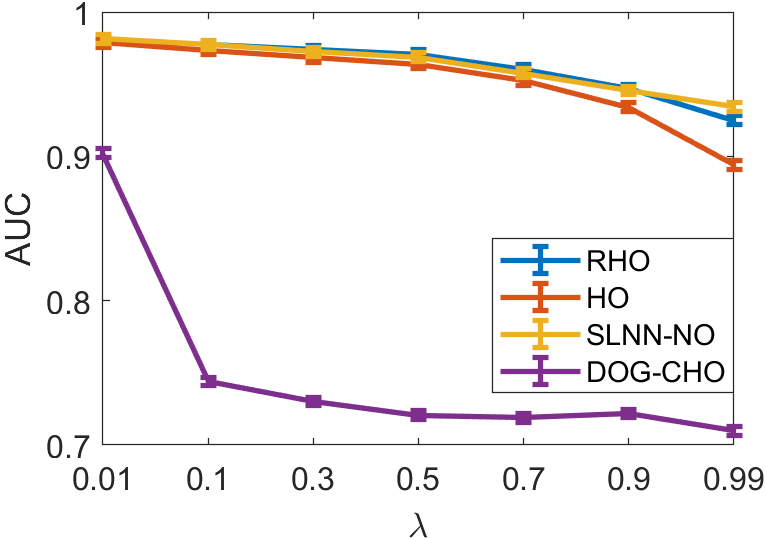}
% \hspace{0.1in}
	\includegraphics[width=0.49\textwidth]{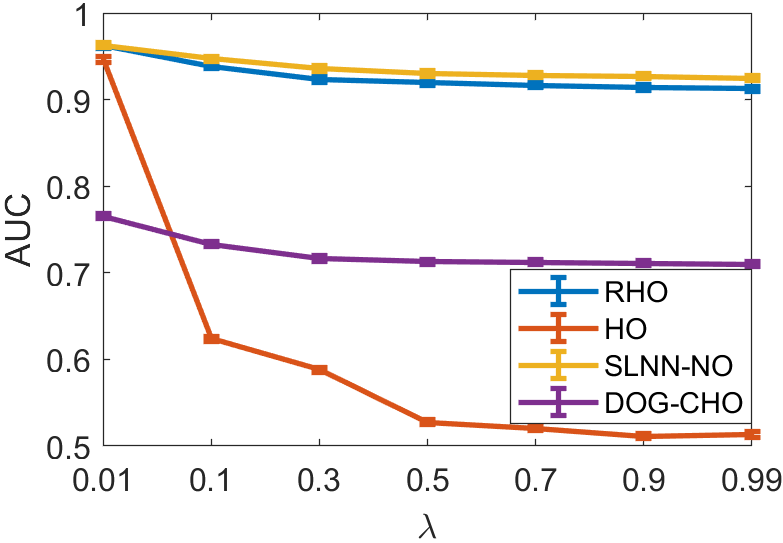}
	\caption{The relationships between AUC and the weight parameter $\lambda$ when different NOs were employed {for objective image quality assessment} were quantified. Both SLNN-HO (top figure) and SLNN-NO (bottom figure) were employed to {compute task component $\mathcal{L}_{\text{t}}$.}
 }
 \vspace{-0.1in}
\label{fig:lknown_lambda}
\end{figure}

The impact of the weight parameter $\lambda$ in Eqn. (\ref{eq:hloss}) on signal detection performance as measured by AUC is shown in Fig. \ref{fig:lknown_lambda}.
Both the SLNN-NO and SLNN-HO described in Sec. \ref{sssec:appended_NO} were considered as the NO to compute task component $\mathcal{L}_{\text{t}}$ in Eqn. (\ref{eq:hloss}).
Signal detection performance was evaluated by {use of} the SLNN-NO, HO,
RHO, and DOG-CHO {acting on the denoised images.}
For both cases, the performance of {the} four different NOs on the denoised images was higher when smaller $\lambda$ values (larger weight for task-based loss) were considered. 
% Larger $\lambda$ represents heavier weight on physical loss for model training.
Those results confirm that the task-informed training method can improve the NOs performance even {when} the NOs {employed for objective image quality assessment} were different from the NO used to compute $\mathcal{L}_{\text{t}}$ {during model training}.
% Even though for the situations where observers for training and evaluation were mismatched, such as SLNN-HO for training and other NOs for evaluation, 
% the task-informed method can still benefit task performance. 
% In addition, for both cases, the performance of RHO was statistically equivalent to that of SLNN-NO, and higher than those of the HO and DOG-CHO.
% The low performance of HO indicates that the corvariance matrices of denoised images were relatively ill-conditioned.

Figure \ref{fig:lknown_lambda} yields two additional noteworthy findings.
%Fourth, As shown in Fig. \ref{fig:lknown_lambda}, 
% First, for the case where the SLNN-HO was employed for training (top figure), the RHO performance was relatively close to that of the HO. However, this relationship can only be generalized to the case where SLNN-NO (bottom figure) was employed with very small $\lambda$ values (e.g., 0.01).
First, for the case where the SLNN-HO was employed for training (top figure), the performance of the HO employed for objective image quality assessment was relatively high (statistically equivalent to that of the SLNN-NO and RHO). However, this relatively high HO performance was only observed for very small $\lambda$ values (e.g., 0.01) in the case where SLNN-NO (bottom figure) was employed.
The HO performance was much lower when relatively large $\lambda$ values (i.e., 0.1-0.99) were employed.
The RHO performance was employed as a reference and was relatively high for all cases.
%When relatively large $\lambda$ values (i.e., 0.1-0.99) were employed, the HO performance was much lower than that of RHO.
These observations suggest that for the case where SLNN-HO was employed for training, the second- and potentially higher-order statistical properties of the images {were} optimized to benefit the HO performance while such behavior {did not} occur in the case where SLNN-NO was considered with large $\lambda$ values.
% The singular value analysis shown in Fig. \ref{fig:lknown_spectra} also confirmed the suggestion.
%Also, As shown in Fig. \ref{fig:lknown_lambda}, 
Second, when SLNN-HO was used for training, the performance of DOG-CHO was greatly improved for small $\lambda$ values and was not significantly improved for other cases. 
This observation indicates that the DOG channels were "closer" to efficient channels when $\lambda$ was appropriately selected.
%This observation indicates that when SLNN-HO was used for task-informed training, the DOG channels were "closer" to efficient channels when $\lambda$ is appropriately selected. 
However, {this behavior was not observed in the case} where SLNN-NO was employed for training.

\subsubsection{Changes in covariance matrix induced by task-informed training method}

\begin{figure}
	\centering
	\includegraphics[width=0.46\textwidth]{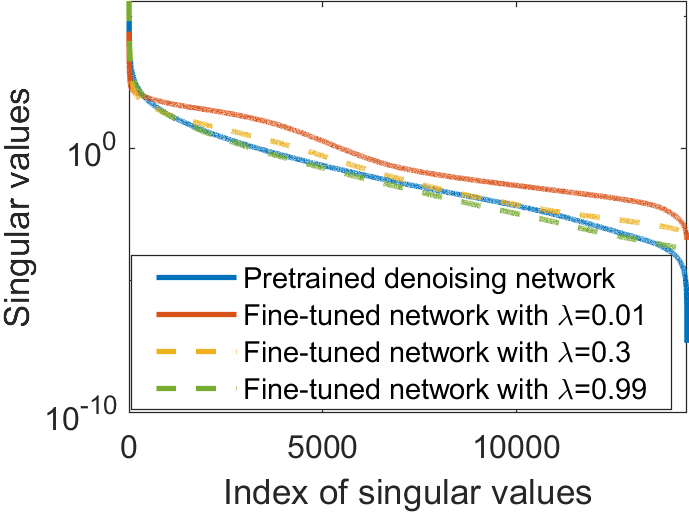}
% \hspace{0.1in}
	\includegraphics[width=0.46\textwidth]{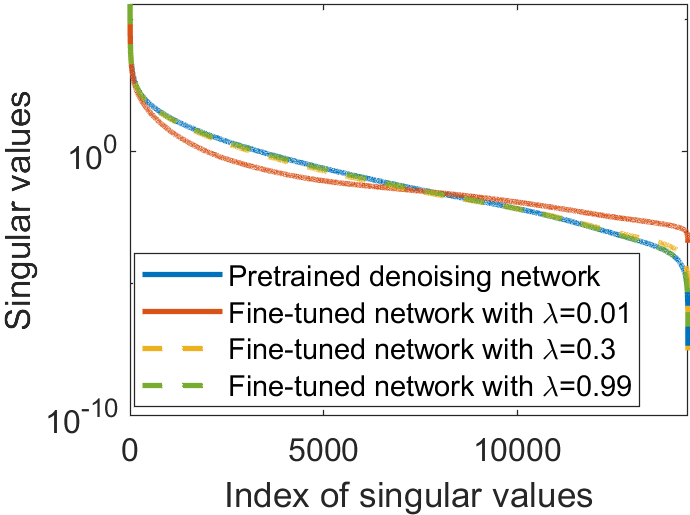}
	\caption{The singular value spectra of covariance matrices
corresponding to the images produced by methods with different $\lambda$. Both SLNN-HO (top figure) and SLNN-NO (bottom figure) were employed to {compute task component $\mathcal{L}_{\text{t}}$.}
 }
 \vspace{-0.1in}
\label{fig:lknown_spectra}
\end{figure}

% The singular value analysis shown in Fig. \ref{fig:lknown_spectra} also confirmed the suggestion.
To gain insights into the behavior of HO performance, the singular value spectra of the covariance matrices corresponding to the images denoised by task-informed training method were further examined. 
The results, shown in Fig. \ref{fig:lknown_spectra}, reveal that the covariance matrix corresponding to the denoised images produced by use of the pretrained denoising network was ill-conditioned, while that corresponding to the denoised images produced by use of the fine-tuned denoising network was well-conditioned when SLNN-HO was employed to compute the task-component $\mathcal{L}_{\text{t}}$ in Eqn. (\ref{eq:hloss}). However, for the case where SLNN-NO was employed to compute $\mathcal{L}_{\text{t}}$, a similar observation only occurred for very small $\lambda$ values (e.g., 0.01) in Eqn. (\ref{eq:hloss}), and the covariance matrices were still ill-conditioned for large $\lambda$ values (bottom figure). 
The results of this analysis {were} consistent with {the} previously discussed results shown in Fig. \ref{fig:lknown_lambda}, and indicated that the task-informed training method may improve the image statistics that are important for signal detection.

\subsection{Results for the case with random signal locations}

\subsubsection{Impact of the weight parameter $\lambda$ and number of trainable layers}

\begin{figure}
	\centering
	\includegraphics[width=0.47\textwidth]{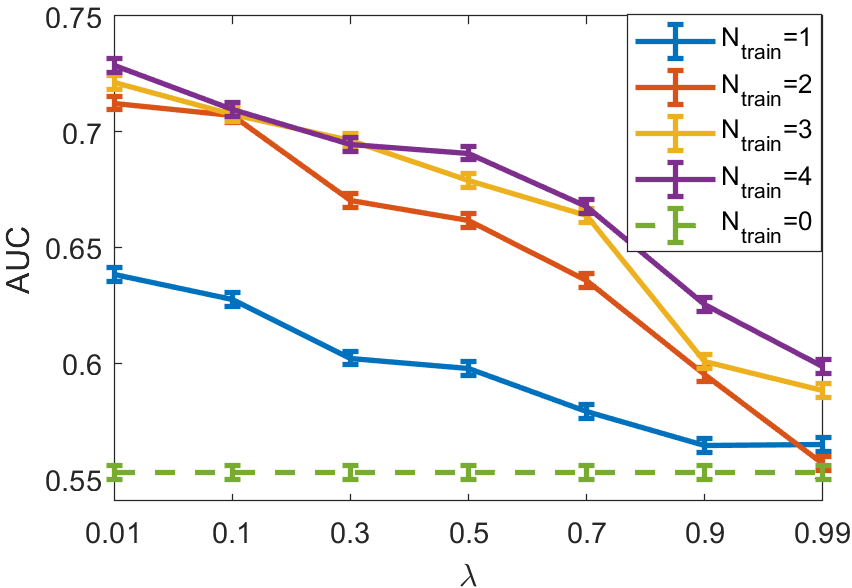}
	\caption{The relationships between AUC and the weight parameter $\lambda$ in the hybrid loss when different numbers of trainable layers are considered were quantified. 
	% The AUC values of the case without fine-tuning were used as references. 
    The quantity $N_{\text{train}}$ means the number of trainable layers of a denoising network. The range of the x-axis is 0.01-0.99.
	The dashed line at the bottom of the figure represents the SLNN-NO performance of the pre-trained denoising network.
	}
	\label{fig:lrandom_lambda}
\end{figure}

\begin{figure}
	\centering
	\includegraphics[width=0.47\textwidth]{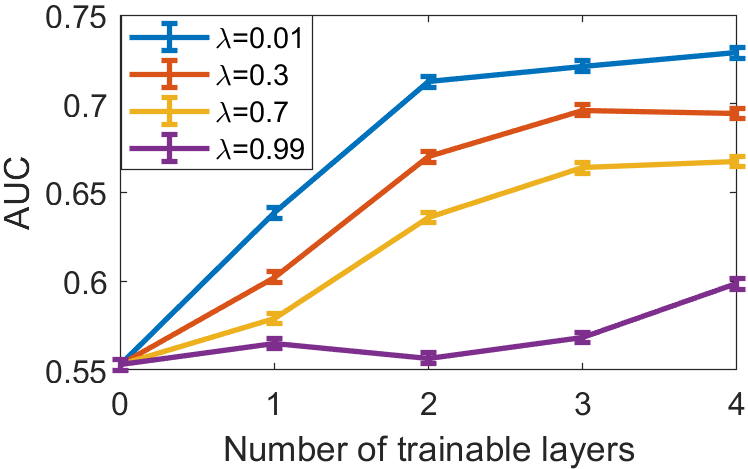}
	\caption{The relationships between AUC and the number of trainable layers of the denoising network. Here the SLNN-NO was employed {to compute task component $\mathcal{L}_{\text{t}}$.} 
 % The task performance was improved after the task-informed training while the traditional IQ metrics were degraded. The improvement of task performance became insignificant when the number of trainable layers and the weight parameter $\lambda$ increased. 
 The trainable layers involved in the fine-tuning started from the last layer and gradually included up to 4 layers.}
 \vspace{-0.1in}
	\label{fig:lrandom_trainable}
\end{figure}

The impacts of the weight parameter $\lambda$ in Eqn. (\ref{eq:hloss}) and the number of trainable layers on {signal detection performance as measured by the SLNN-NO} are {shown} in Figs.~\ref{fig:lrandom_lambda} and \ref{fig:lrandom_trainable}, respectively. 
Here the SLNN-NO employed to compute the task-component $\mathcal{L}_{\text{t}}$ in Eqn. (\ref{eq:hloss}) was also employed {to assess signal detection performance.}
For comparison, the impact on traditional measures of IQ is demonstrated in Table \ref{tab:CNN_RMSE}. 
It was observed that, after the task-informed model training, the SLNN-NO signal detection performance was improved while the traditional measures of IQ were degraded compared with those achieved by the pre-trained denoising network.

As shown in Fig.~\ref{fig:lrandom_lambda} and Table \ref{tab:CNN_RMSE}, for all numbers of trainable layers, the task performance increased while $\lambda$ decreased. 
%while the traditional measures of IQ were degraded.
In addition, the degradation of traditional metrics was significant for relatively small $\lambda$ 
while insignificant for large $\lambda$ values (i.e., $\lambda=0.3-0.99$).
As expected, the trade off between traditional and task-based measures of IQ can be controlled by $\lambda$.
For example, when $\lambda=0.3$, the resulted AUC value was greatly improved to that of $\lambda=0.99$, 
while the RMSE and SSIM were statistically equivalent to that of $\lambda=0.99$.

% It was observed that, the improvement of traditional metrics were insignificant for relatively large $\lambda$ values (i.e., $\lambda=0.3-0.99$).
% \textcolor{blue}{Thus, for example, when $\lambda=0.3$, the resulted AUC value was slightly smaller than that in the case when $\lambda=0.01$, while the RMSE and SSIM were statistically equivalent to those in the case when $\lambda=0.99$.}

As shown in Fig.~\ref{fig:lrandom_trainable}, for relatively small $\lambda$ (i.e. where the task-based component is highly weighted),
significant improvements in task performance were achieved by fine-tuning the last (or last several) convolutional layer(s) (e.g., 0-2) while improvement was insignificant when more layers were trainable.
In addition, when larger values (i.e., $\lambda=0.99$) were selected, the improvement in task performance was
irrelevant to the number of trainable layers. 
% statistically insignificant when only a few layers (e.g., 0-3) were trainable. 
For traditional IQ metrics (i.e., RMSE, SSIM), Table \ref{tab:CNN_RMSE} showed that the degradation mainly resulted from the last small number of convolutional layer(s). 
For larger $\lambda$ values (i.e., $\lambda=0.7\sim0.99$), the changes in traditional IQ metrics were statistically insignificant.

\begin{figure}[h]
     \centering
     \includegraphics[width=0.45\textwidth]{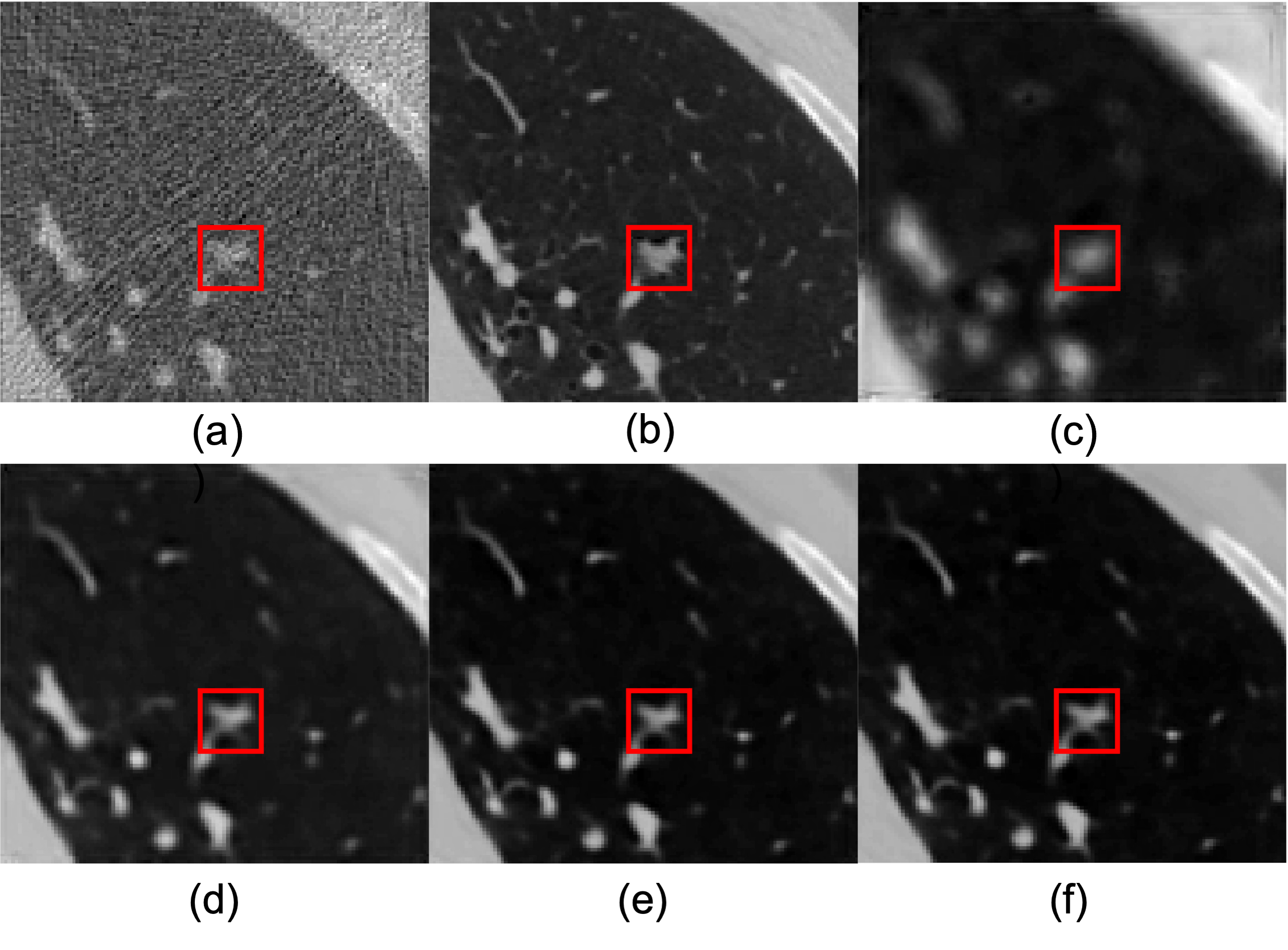}
    \caption{The images represent (a) a low-dose signal-present image, (b) a normal-dose signal-present image, and (c)-(f) denoised estimates $\hat{\mathbf{f}}$ generated by the task-informed denoising method with $\lambda=0.01, 0.1, 0.5, 0.99$, respectively.
    The red box indicates the inserted signal.
    }
    \vspace{-0.1in}
    \label{fig:visual_lrandom}
\end{figure}

The denoised estimates produced by the task-informed image denoising methods were also subjectively assessed. Figure \ref{fig:visual_lrandom} shows a noisy image and denoised images generated by denoising networks fine-tuned with the hybrid loss for different values of  $\lambda$ in Eqn. (\ref{eq:hloss}).
%As shown in Fig. \ref{fig:visual_lrandom}, 
The denoised estimates were blurred as a result of the task-informed training, and the level of blur increased when $\lambda$ decreased. 

\begin{table}
	\centering
	\small
	\caption{The relationships between RMSE and the number of trainable layers of the denoising network and the weight parameter $\lambda$. The traditional IQ metrics were degraded after the task-informed training {and the degradation mainly resulted from the last small number of convolutional layer(s).} The quantity $N_{\text{train}}$ represents the number of trainable layers.
	%of a CNN-based denoising network during the fine-tuning.
	}
	\begin{tabular}{c|c| c| c |c | c | c}
    \hline\hline
    $N_{train}$ &$\lambda$ & 0 & 1 & 2 & 3 & 4\\
    \hline
     \multirow{7}{*}{RMSE}     &0.01   & 1.3437 &1.7178 & 1.9304 & 2.1127 & 2.2595   \\ 
     &0.1   & 1.3437 &1.3775 & 1.3988 & 1.4147 & 1.4244   \\
    &0.3   & 1.3437 &1.3509 & 1.3549 & 1.3613 & 1.3667   \\
    &0.5   & 1.3437 &1.3506 & 1.3514 & 1.3534 & 1.3546   \\
    &0.7   & 1.3437 &1.3507 & 1.3506 & 1.3501 & 1.3498   \\
    &0.9   & 1.3437 &1.3503 & 1.3499 & 1.3498 & 1.3495   \\ 
    &0.99   & 1.3437 &1.3498 & 1.3485 & 1.3481 & 1.3477   \\ 
    \hline
    \multirow{7}{*}{SSIM}     &0.01   & 0.9416 &0.9197 & 0.9024 & 0.8755 & 0.8738   \\
    &0.1   & 0.9416 &0.9391 & 0.9381 & 0.9369 & 0.9368   \\
    &0.3   & 0.9416 &0.9404 & 0.9401 & 0.9394 & 0.9391   \\ 
    &0.5   & 0.9416 &0.9405 & 0.9407 & 0.9407 & 0.9408   \\ 
    &0.7   & 0.9416 &0.9407 & 0.9408 & 0.9409 & 0.9409   \\ 
    &0.9   & 0.9416 &0.9407 & 0.9409 & 0.9411 & 0.9412   \\
    &0.99   & 0.9416 & 0.9408 &0.9409 & 0.9410 & 0.9414   \\ 
	\hline\hline
	\end{tabular}
 % \vspace{-0.05in}
	\label{tab:CNN_RMSE}
\end{table}

% \subsubsection{Visual analysis of the task-informed training methods}

% Figure \ref{fig:visual_lrandom} shows a noisy image and denoised images generated by denoising networks fine-tuned with the hybrid loss for different values of  $\lambda$ in Eqn. (\ref{eq:hloss}).
% %As shown in Fig. \ref{fig:visual_lrandom}, 
% The denoising estimates were blurred after the task-informed training, and the level of blur increased when $\lambda$ decreased. 
% To investigate the relationship between the level of blur and the improvement in task performance,
% line profile analysis was performed
% to analyze the intensity variation crossing a nodule or nodule-like structure. 
% % As shown in Fig. \ref{fig:lrandom_visual}, the distance between profiles from denoised estimates with $\lambda=0.99$ and profiles from the normal-dose images was smaller than that calculated from denoised estimates with $\lambda=0.01$. 
% As shown in Fig. \ref{fig:lrandom_visual}, the variations of profiles from denoised estimates with $\lambda=0.01$ were more similar to those from the normal-dose images while was smoothed out in the case of $\lambda=0.99$.
% % This indicated the nodule-like structures from the denoised estimates with $\lambda=0.99$ were more homogeneous. 
% The observations validated that increasing the $\lambda$ will improve the physical measures while compromising some structure details and also revealed the potential explanation for the improvement in task performance.

% \begin{figure}[h]
%      \centering
%      \includegraphics[width=0.5\textwidth]{figure/lambda01_trainable.png}
%     \caption{The images represent (a) a low-dose signal-present image, (b) a normal-dose signal-present image, and (c)-(f) denoised estimates $\hat{\mathbf{f}}$ generated by the task-informed training method with last \{1, 2, 3, 4\} layers trainable, respectively.
%     }
%     \label{fig:sample}
% \end{figure}

% \begin{figure}
% 	\centering
% 	\includegraphics[width=0.5\textwidth]{figure/lrandom_visual.png}
% 	\caption{The visual analysis of the denoised images. The images, from left to right, in each row represent (a) the normal dose image, (b) the denoised estimates with $\lambda=0.01$ and (c) $\lambda=0.99$, and (d) the profiles of the indicated line in the figures. Each curve in Fig. (c) represents the profile corresponding to the line in Figs. (a)-(c) with the same color. The variations of the profile from the denoised image with $\lambda=0.01$ were similar to those from the normal dose.
% 	}
% 	\label{fig:lrandom_visual}
% \end{figure}

\subsubsection{Impact of denoising network depth}
A study was performed to investigate whether the loss of task-relevant information {primarily} occurs in the last several layers when the denoising network depths increase. As shown in Table \ref{tab:lrandom_depth}, 
%for the pretrained denoising networks (trained using traditional loss alone), 
the SLNN-NO performance decreased as a function of denoising network depth, which is consistent with previous findings \cite{li2021assessing} that the mantra "deep is better" may not always hold true for objective IQ measures. After the task-informed training, the SLNN-NO performance was improved and the variations of the improved SLNN-NO performance were statistically insignificant when network depth varied. No matter how deep the pretrained denoising network was, the loss of task-relevant information still occurred in the last certain layers (i.e., not related to the depth of denoising networks), at least in the considered cases.

\begin{table}
	\centering
	\small
	\caption{The relationship between the SLNN-NO performance and the depth $D$ of the denoising networks. Only the last three layers were trainable for fine-tuning. The standard error for AUC values is 0.003.}
	\begin{tabular}{c|c| c| c |c }
    \hline\hline
    $D$ & 9 & 11 & 13 & 15\\
    \hline
    AUC (Fine-tuned)    &0.6984   & 0.7146 & 0.7074 & 0.7130   \\ 
    AUC (Pretrained) &0.5751   & 0.5714 &0.5529 & 0.5501   \\

	\hline\hline
	\end{tabular}
 % \vspace{-0.1in}
	\label{tab:lrandom_depth}
\end{table}

% \begin{figure}
% 	\centering
% 	\includegraphics[width=0.45\textwidth]{figure/auc_denoisingdepth.png}
% 	\caption{The visual analysis of the denoised images. The images, from left to right, in each row represent normal dose image, the denoised estimates with $\lambda=0.01$ and $\lambda=0.99$, and the profiles of the red dashed line in the figures. The variations of the profile from the denoised image with $\lambda=0.01$ were similar to those from the normal dose.
% 	}
% 	\label{fig:lrandom_depth}
% \end{figure}

\subsection{Impact of the task shifts for training and evaluation}
\begin{figure}
	\centering
	\includegraphics[width=0.47\textwidth]{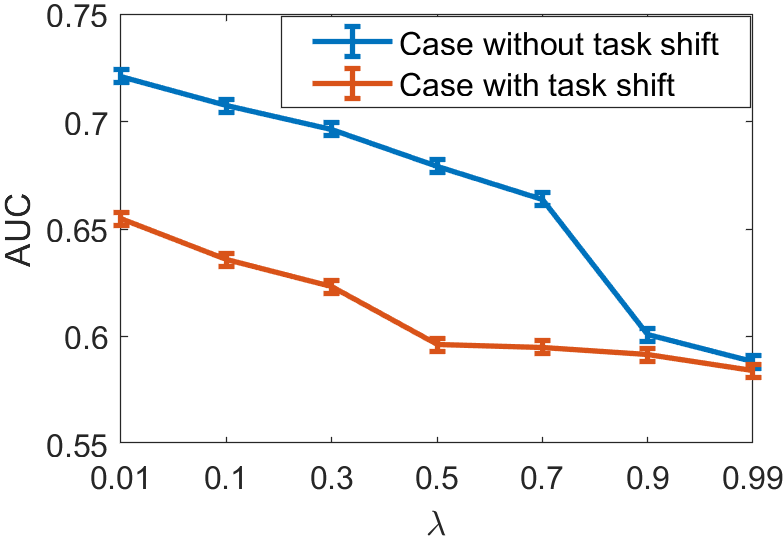}
% \hspace{0.1in}
	\includegraphics[width=0.47\textwidth]{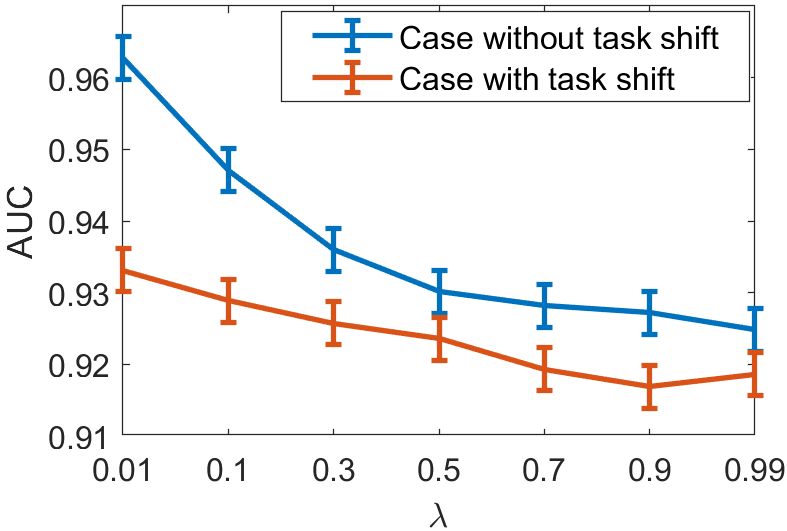}
	\caption{The performance of the SLNN-NO for the case with and without task shift was compared. Two cases were considered: (1) binary signal detection tasks with fixed signal locations were employed for training and the tasks with random signal locations were employed for evaluation (top figure); (2) the tasks with random signal locations were employed for training and the tasks with fixed signal locations were employed for evaluation (bottom figure).
 }
 \vspace{-0.1in}
\label{fig:task_shift}
\end{figure}

% \begin{figure}
% 	\centering
% 	\includegraphics[width=0.48\textwidth]{figure/ALROC_4_train.png}
% % \hspace{0.1in}
% 	\includegraphics[width=0.48\textwidth]{figure/ALROC_2_train.png}
% 	\caption{The performance of the SLNN-NO for the case with and without task shift was compared with different $\lambda$ values. Two cases were considered: (1) detection-localization tasks with 2 possible signal locations were employed for training and the tasks with 4 possible signal locations were employed for evaluation (top figure); (2) detection-localization tasks with 4 possible signal locations were employed for training and the tasks with 2 possible signal locations were employed for evaluation (bottom figure).
%  }
% \label{fig:task_shift_lroc}
% \end{figure}
The robustness of the task-informed image denoising method to task-shift was also assessed and the results are shown in Fig. \ref{fig:task_shift}.
% The impact of the shift in tasks for training and evaluation was also investigated and the results are shown in Fig. \ref{fig:task_shift}. 
The SLNN-NO performance for the case with no task shift was considered as a reference. {It was observed that introducing task shift always degraded the task performance as expected and the degradation} resulting from task shift became insignificant when $\lambda$ increased. This is due to the smaller weight for the task-based loss component as $\lambda$ increases, which makes the impact of task shift less significant. For the case where a relatively simple task was used for training and the complex one was used for evaluation (top figure in Fig. \ref{fig:task_shift}), the degradation in task performance was much more significant than in the case where a complex task was used for training and a simple one was used for evaluation (bottom figure in Fig. \ref{fig:task_shift}). 
% This observation is due to the fact that the case with fixed signal locations can be considered as a special case to the case with random signal locations. 
This observation is due to the fact that the case with random signal locations can be easily generalized to the case with fixed signal locations but {not} vice versa. 
% Similar observations can also be generalized to detection-localization tasks with both 2 and 4 possible signal locations considered. The relatively simple task (2 possible signal locations) can be considered as a special case to the relatively complex task (4 possible signal locations).
% This observation suggested that employing a relatively complex task other than a simple task can improve the robustness of a task-informed image restoration method to task-shift.
This suggested that employing a relatively complex task for training can better improve the robustness of a task-informed image restoration method to task-shift than employing a simple task.

% \subsection{Impact of the data shifts for training and evaluation}
% The impact of the shift in data for training and evaluation is shown in Fig. \ref{fig:data_shift}. The SKS/BKS binary signal detection tasks with fixed signal locations were considered. The shift in data was created by changing the noise distribution that was described in Sec. XXX. The SLNN-NO performance of the case with no shift in the data was considered as a reference case. It was observed that, for each $\lambda$ value, the shifts in data will degrade the task performance as expected. Specifically, for both cases, the degradation resulting from data shift became insignificant when $\lambda$ increased considering that the the task-based loss component has smaller weight as $\lambda$ increases, which makes the impact of data shift less significant.

% \begin{figure}
% 	\centering
% 	\includegraphics[width=0.45\textwidth]{figure/data_shift.png}
% 	\caption{The performance of the SLNN-NO for the case with and without data shift was compared with different $\lambda$ values. The SKS/BKS binary signal detection tasks with deterministic signal locations were considered.
% 	}
% 	\label{fig:data_shift}
% \end{figure}

\section{Discussion and Summary}
\label{sec:summary}

In this work, a task-informed DNN-based image denoising method that preserves task-specific information was {objectively} evaluated. 
This study was motivated by previous works \cite{li2021assessing,yu2020ai} that indicate traditional DNN-based denoising methods may not benefit task performance even though the traditional measures of IQ were improved.
% It should be noted that although the considered tasks are not as realistic as clinical situation, the designed method can still be investigated and studied corresponding behaviors.
The task-informed model training method employed a hybrid loss strategy and only acted on the last several layers of a DNN-based denoising method. To evaluate the method, binary signal detection tasks with fixed and random signal locations under SKS/BKS conditions were considered.
%Both fixed and random signal locations were considered to enable assessment by use of NOs and demonstrate the ability to generalize to complicated cases. 
The performance of SLNN-NO, SLNN-HO, and common NOs was quantified to assess the impact of task-informed training on task performance preservation.

The numerical results indicated that, in the cases considered, the task-informed training method could preserve task-relevant information in the image and control the trade off between traditional and task-based measures of IQ. 
% by the selection of $\lambda$ and the number of layers in the denoising network that are subject to fine-tuning.
The performance of the task-informed training method was influenced by the weight parameter $\lambda$, the number of trainable layers, the depth of denoising network, and the task complexity.
% The results indicated that the trade off between traditional and task-based measures of IQ can be controlled by the selection of $\lambda$ and the number of layers in the denoising network that are subject to fine-tuning.
The evaluation also confirmed that traditional IQ metrics may not always correlate with task-based metrics. 
Additional insights were gained by performing singular value spectra analysis. 
It was revealed that the method could mitigate the ill-conditioning of covariance matrices and have the potential to improve the image statistics that are important for signal detection.

To better understand the potential suitability of a task-informed image restoration method for clinical translation, its robustness to task-shift was also assessed.  It was observed that introducing task-shift will degrade the task performance as expected. The degradation was significant when a relatively simple task was considered as source task while a complex one was used as target task. The degradation can be potentially mitigated by employing the complex task as source task and the simple one as target task. 
% The results were also confirmed by the case with detection-localization tasks that are described in the supplemental file. 
% The results suggested that employing a relatively complex task {other than a simple task} can improve the robustness of a task-informed image restoration method to task-shift and subsequently accelerate the clinical translation. 
{This suggests} that employing a relatively complex task for training can better improve the robustness of a task-informed image restoration method to task-shift than employing a simple task.

There remain numerous important topics for future investigation. 
In this work, the SLNN-NO and SLNN-HO were employed to compute the task-component $\mathcal{L}_{\text{t}}$ in Eqn. (\ref{eq:hloss}). Anthropomorphic numerical observers (ANOs) may {instead} be employed to predict human observer performance \cite{abbey2001human,zhang2006effect,han2020convolutional}. 
Employing an ANO to compute $\mathcal{L}_{\text{t}}$ may 
potentially benefit a task-informed image denoising method if humans are the ultimate readers of the image.
The evaluation study in this paper focuses on significant parameters such as the weight parameter $\lambda$ in Eqn. (\ref{eq:hloss}) and the number of trainable layers. 
Other parameters, such as the size of training dataset and the ratio between signal-present and signal-absent images, remain unexplored.
%size of training dataset, the ratio between signal-present and signal-absent images., 
% The {task-informed image denoising method} needs an exactly defined task.
% However, the knowledge of the exactly defined task is unavailable in real situations. It will be important to design a method that is robust when tasks vary.
% \textcolor{blue}{It should be noted that the considered tasks are not as realistic as clinical situation}.
The extension of the proposed method for use with more complex tasks such as detection-estimation tasks \cite{li2021hybrid, clarkson2007estimation} is also an important topic.
The {task-informed image denoising method} and the corresponding assessment {strategy} can also readily be applied to different image restoration and reconstruction methods. 
Ultimately, it will be critical to conduct human reader studies to assess the benefit of {any} task-informed method.

% \vspace{-0.1in}
\bibliography{task-denoise-spie} % bibliography data in report.bib
\bibliographystyle{IEEETran}

\end{document}